\documentclass[11pt,letterpaper,showpacs,superscriptaddress]{article}
\usepackage{graphicx}
\usepackage{psfrag}
\usepackage{color}
\usepackage{amsbsy}
\usepackage{amssymb}       

\def\a{\alpha}
\def\b{\beta}

\def\s{\sigma}

\begin{document}


\title{Clustered bottlenecks  in mRNA translation and protein synthesis}

\author{Tom Chou$^{1}$ and Greg Lakatos$^{2}$ \\
$^{1}$Dept. of Biomathematics and IPAM, 
UCLA, Los Angeles, CA 90095 \\ $^{2}$Dept. of Physics, University of British Columbia, 
Vancouver, BC Canada V6T 1Z2}

\date{\today}

\maketitle 

\begin{abstract}

We construct  an algorithm that generates large, band-diagonal
transition matrices for a totally asymmetric exclusion process
(TASEP) with local hopping rate inhomogeneities.  The matrices
are diagonalized numerically to find steady-state currents of 
TASEPs with local variations in hopping rate. 
The results are then used to investigate
clustering of slow codons along mRNA. Ribosome density
profiles near neighboring clusters of slow codons interact,
enhancing suppression of ribosome throughput when such
bottlenecks are closely spaced. Increasing the slow codon cluster
size, beyond $\approx 3-4$, does not significantly reduce ribosome current.
Our results are verified by extensive Monte-Carlo simulations
and provide a biologically-motivated explanation for the
experimentally-observed clustering of low-usage codons.  
\end{abstract}

\noindent 05.10.-a, 87.16.Ac, 87.10.+e

\vspace{7mm}


\baselineskip=14pt

\begin{figure}[h]
\begin{center}
\includegraphics[height=2.25in]{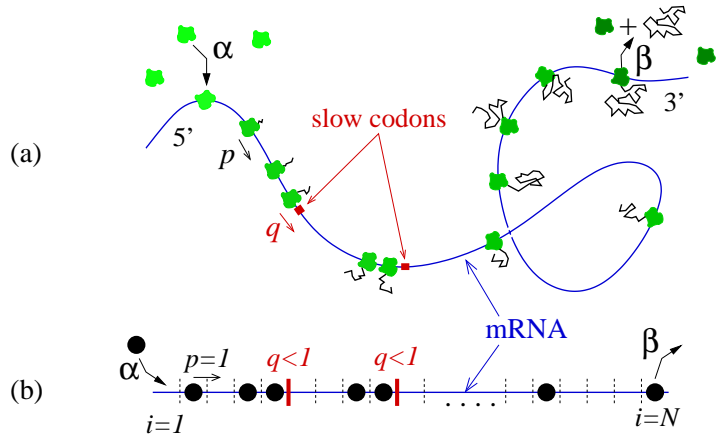}
\end{center} 
\vspace{-4mm}
\caption{(a) The mRNA translation/protein synthesis
process. Multiple ribosomes (polysomes) move unidirectionally along
mRNA as tRNA (not shown) deliver the appropriate amino acid to the
growing chain.  Codons with low concentrations of corresponding
tRNA result in bottlenecks that locally suppress
ribosome motion across it. (b) A simple totally asymmetric
exclusion process on $N$ lattice sites used to model mRNA
translation in the presence of slow codons bottlenecks, or ``defects.''} 
\label{MRNA} 
\end{figure}

During protein synthesis, ribosome molecules initiate at the 5'
end of messenger RNA (mRNA), scan (``elongate'') along the
mRNA sequence, and terminate with the completed protein
product at the 3' termination end (Fig.  \ref{MRNA}a).  Each
elongation step requires reading (translating) a nucleotide triplet
(codon) and the binding of a freely diffusing transfer RNA (tRNA)
molecule carrying the amino acid specific to each codon
\cite{MBOC}.  Besides being a critical final stage of gene
expression {\it in vivo}, control of protein synthesis is vital protein
adaptation and evolution
\cite{BULMER2,DURET,MORIYAMA}, in control of
viral parasitism \cite{KURLAND}, and in the biotechnology of
high yield, cell-free, synthetic {\it in vitro} protein production
\cite{UEDA,CELLFREE}.  

The near-final stage of protein expression can be regulated by
exploiting local variations in ribosome detachment and
elongation rates (arising from different codon usages for any
given amino acid).  Relative concentrations of tRNA in solution
can  be a factor in determining  local ribosome translation rates. 
This fact has been exploited in the modification of green
fluorescent protein codons to match those preferred in mammalian systems,
thereby optimizing expression levels \cite{TSIEN}.  Binding of
an incorrect tRNA can also  temporarily prevent binding of the
correct tRNA for a particular codon, slowing down elongation
\cite{MBOC}. Regions of higher ribosome density can also
protect the substrate mRNA from the action of enzymes which
arrest translation.  Finally, local pauses  may  provide time for
regions of the peptide to sequentially fold before the integration
of additional amino acids.




The occurrence of ``slow''  codons (those with rare corresponding tRNA and/or
amino acids), along mRNA is known to inhibit protein production
\cite{ROBINSON,SORENSEN}, and  is conjectured to serve a regulatory role in
protein synthesis. Such  ``bottleneck'' or ``defect''  codons (and the amino acids
for which they code) arise infrequently, and typically include CTA (Leu), ATA
(Ile), ACA (Thr), CCT and CCC (Pro), CGG, AGA, and AGG (Arg)
\cite{CLUSTER2,GODSON}.  Statistics indicate a higher occurrence of rare
codons near the 5' initiation site of {\it E.  coli} genes \cite{CHEN}. Even more
striking is the bias for rare codons to cluster \cite{CLUSTER2}.  Statistical
analyses show small (2-5) clusters of rare codons occur frequently in {\it E. coli},
{\it Drosophilia}, yeast, and primates \cite{CLUSTER2,CLUSTER1}.  

Although the strength, number, and positioning of bottlenecks
along an mRNA chain undergoing translation clearly  affect local
ribosome densities and overall translation rates, there has been no
quantitative model describing how various bottleneck motifs
control ribosome throughput and protein synthesis.  
In this paper, we consider a simple physical model of how specific codon
usages that give rise to local delays in elongation can be used to
suppress protein synthesis.  Our results provide biophysical
hypotheses for slow codon clustering within a simple
nonequilibrium steady-state stochastic framework.  Ribosome
density boundary layers surrounding two nearby bottlenecks, or
defects are shown to act synergistically to regulate and amplify
the suppression of protein production.

We model mRNA translation by ribosome particles using a
nonequilibrium  totally asymmetric exclusion process (TASEP)
\cite{DER98,SHAW} with a few carefully distributed slow
sites (cf. Fig.  \ref{MRNA}b).  In the TASEP, ribosome particles 
attach (initiate) at the first lattice site with rate $\alpha$, only if
the first site is empty.  Interior ribosome particles can move
forward with rate $p_{i}$ from site $i$ to site $i+1$ only if site
$i+1$ is empty.  For each step a ribosome moves forward, a
codon is read, and an amino acid-carrying tRNA delivers its
amino acid to the growing polypeptide chain.  No motion is
allowed if the site in front of a particular ribosome is occupied. 
Each ribosome that reaches the last site $i=N$ (the 3' termination
site) has polymerized a complete protein and detaches with rate
$\beta$.  

In the case of {\it uniform} $p_{i} = 1$, the protein production rate
({\it e.g.} the steady-state ribosome current)  and ribosome
density along the mRNA are known exactly in terms of $\alpha$
and $\beta$ \cite{DER98,SHAW}.  In the long chain
($N\rightarrow \infty$) limit, the steady-state results reduce to simple
forms illustrating the fundamental physical regimes.  The current
may be entry-rate limited ($\a < 1/2$, $\a < \b$), where the
ribosome density is low and the steady-state current
$J(\a)=\a(1-\a)$ depends only on $\a$.  If $\beta$ is sufficiently
small ($\b<1/2, \b<\a$), the density is high and the current
$J(\b)=\b(1-\b)$ is a function of the rate-limiting exit step only. 
When both $\a, \b > 1/2$, the rate-limiting processes are the
uniform internal hopping rates, and the steady-state current
reaches its maximum attainable value $J=1/4$.  
These three regimes define a phase diagram or the 
steady-state current of the TASEP. 
For a typical mRNA length of $\sim 100-1000$ codons,
these simple analytic forms for $J$ are extremely accurate.  For
the analyses in the absence of ribosome detachments, 
we shall restrict our attention to
effectively the $N\rightarrow \infty$ limit. 

\begin{figure}[h]
\begin{center}
\includegraphics[height=1.0in]{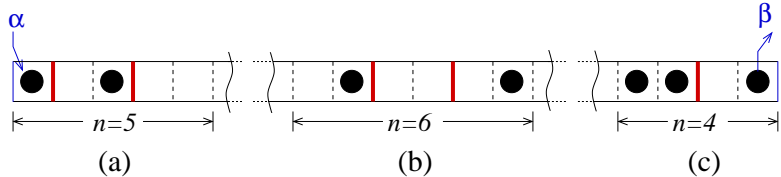}
\end{center}
\vspace{-4mm}
\caption{Placements of slow defects, or rare-usage codons 
(thick red segments).
(a) Two defects near the initiation site straddled by an $n=5$ lattice segment.
(b) Two defects in the 
chain interior, away from the boundaries, $n=6$. (c) 
A single slow defect near the 
termination end of the chain, $n=4$.}
\label{SITES}
\end{figure}

\begin{figure}
\begin{center}
\includegraphics[height=1.5in]{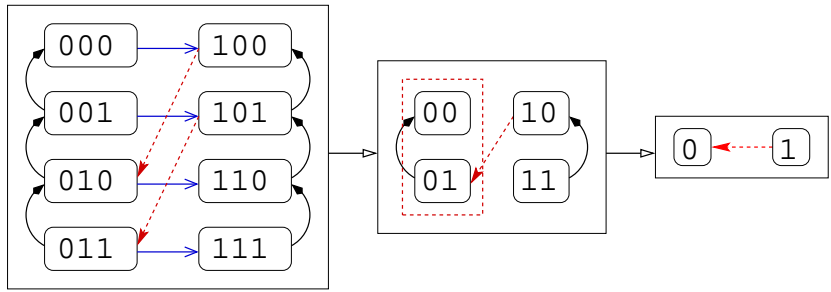}
\end{center}
\vspace{-4mm}
\caption{The algorithm that generates the transition matrix for a
TASEP.  For illustration, consider a three site model.  Each
possible occupancy of the lattice is associated with a bit pattern,
and the state is enumerated with the corresponding decimal
value.  For example given that \texttt{011} is the binary
representation of \texttt{3}, we label the state with particles
occupying the second and third lattice sites as \textit{state 3}. 
Next we divide the states into two groups; states where the first
lattice site is occupied \textit{(1-states)}, and states where the
first lattice site is empty \textit{(0-states)}.  Regardless of the
number of lattice sites in the TASEP, the transitions between the
two classes of states \textit{always} occur between the first half
of the \textit{1-states} and the second half of the \textit{0-states}
(dashed red arrows).  To determine the
remaining transitions we call the algorithm recursively on both
the \textit{1-states} and the \textit{0-states}, making sure to
ignore the highest order bit ({\it i.e.} the leftmost lattice site). 
Finally we add in the trivial transitions between each
\textit{0-state} and each \textit{1-state} that are the result of
injection at the left edge of the lattice (blue arrows).  
With the connectivity of the states fully enumerated,
assigning the appropriate rate $p_i$ to each of the transitions is
straightforward.}
\label{RECURSION}
\end{figure}

There is no general theory for  computing steady-state particle (ribosome)
currents when the internal hopping rates $p_{i}$ vary arbitrarily with lattice
position $i$. However, specific motifs $\{p_{i}\}$ (such as isolated defects
\cite{KOLODEFECT} and periodic variations) can be treated with approximations
and simulation.  Since slow codons are also rare, with typical densities of 0.03,
we will consider simple configurations of few, identical bottlenecks (with hopping
rate $q<1$) distributed within an otherwise uniform (with rate 1) lattice.  Finite
numbers of ``fast''  defects with $q>1$ do not affect steady-state currents since
the rate limiting hops $p_{i} = 1$ dominate the lattice.  Figures \ref{SITES}(a-c)
show hypothetical placements of defects near the 5' initiation end, the mRNA
interior, and the 3' termination end.  Within a finite segment (of length $n=5,6,4$
sites in Figs. \ref{SITES}, respectively) straddling the bottlenecks, we explicitly
enumerate all $2^{n}$ distinct states according to the algorithm indicated in Fig. 
\ref{RECURSION}. This generates a {\it band-diagonal} (of width $n$)
$2^{n}\times 2^{n}$ matrix coupling the probabilities $P_{j},\, (1\leq j \leq 2^{n})$
that the segment is in state $j$.  

Now consider an interior segment (Fig.  \ref{SITES}(b)). The mean
density in the site immediately to the left(right) of the segment is
denoted $\sigma_{-}(\sigma_{+})$. The transition matrix contains the
parameters $\s_{\pm}$ since entry and exit into the enumerated
segment is proportional to $\sigma_{-}$ and $1-\sigma_{+}$,
respectively.  The steady-state current is calculated from the
appropriate elements of the singular eigenvector associated with the
zero-eigenvalue of the transition matrix.  For example, the particle flux
out of the rightmost site of the segment is $J(\s_{-},\s_{+}, \{p_{i}\}) =
(1-\s_{+})\sum_{j=odd}P_{j}(\s_{-},\s_{+}, \{p_{i}\})$, since the odd
states correspond to those with a particle at the last site in the segment. 
The singular eigenvector of the $2^{n}\times 2^{n}$, sparse,
band-diagonal transition matrix was computed (up to $n\approx 18$)
using an implicit restarted Arnoldi iteration method \cite{ARNOLDI}
found in Matlab. The effects of the perturbed particle correlations
surrounding a bottleneck  are accounted for provided $n$ is larger than
the density boundary layer thickness.  Since the densities are uniform
far enough away from the defects, we assume that they  are also
$\sigma_{-}(\sigma_{+})$ far to the left(right) of the segment.  Thus, the
mean-field currents well to the left and right of the segment are $J_{-}
= \s_{-}(1-\s_{-}) = J_{+}=\s_{+}(1-\s_{+})$. The only physical solution
is $\s_{-} = 1-\s_{+}$. Equating $J(\s_{-}=1-\s_{+}, \s_{+}, \{p_{i}\}) =
\s_{+}(1-\s_{+})$, and solving for $\s_{+}$ numerically, we find the
ribosome current $J$.  For $\a<1/2$ and defects near the initiation end,
we simply equate $J(\a,\s_{+}, \{p_{i}\}) = \s_{+}(1-\s_{+})$ and solve for
$\s_{+}$. For $\b<1/2$ and defects near the termination end, $\s_{-}$ is
determined from $\s_{-}(1-\s_{-}) = J(\s_{-}, \beta, \{p_{i}\})$.  We used
this improved, systematic 
finite-segment mean-field theory (FSMFT) to compute
currents of TASEPs with representative placements of internal defects. 
In practice, segment lengths that include only 2-3 sites on each side of a
defect were sufficient for obtaining extremely accurate results.  Efficient
continuous-time Monte-Carlo simulations using the
Bortz-Kalos-Lebowitz (BKL) algorithm \cite{BKL,LAKATOS}
were performed on lattices of size $N\geq 10000$ to verify and extend
all our results.

\begin{figure}[h]
\begin{center}
\includegraphics[height=2.25in]{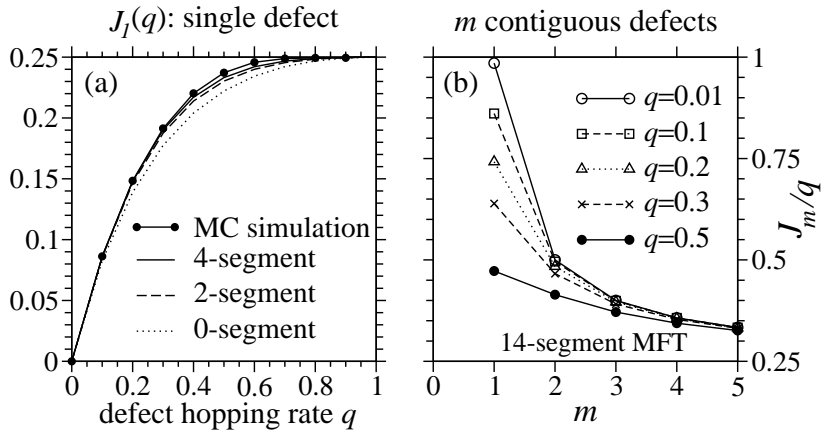}
\end{center}
\vspace{-4mm}
\caption{(a) Comparison of steady-state currents $J_{1}(q)$ for a single defect
derived from MC simulations with those obtained from $n$-segment MFT. 
For $n\geq 4$, the FSMFT results are within $2\%$ of those from MC simulations.
The boundary injection/extraction rates were not rate limiting $(\a=\b=10)$.
(b) Further reduction of steady-state current as successive, 
identical defects are added. The first few defects 
cause most of the reduction in current.}
\label{single_mid2}
\end{figure}

MC simulations show that for $\a,\b>1/2$, the currents $J$ are
insensitive to the position of defects. The behavior for $\a,\b < 1/2$
resembles an interior defect near the initiation or termination end.  Slow
initiation and/or termination rates can be effectively described by
defects ($q<1$) near the ends of the lattice. Therefore, we can restrict
our analysis to large $\a,\b\gg 1/2$.  Hopping across a single interior
defect (with rate $q<1$)  is the overall rate-limiting step. For this single
defect, the reduced steady-state current $J_{1}(q)$, found from both
FSMFT and MC simulations are shown in Fig.  \ref{single_mid2}(a).  The
$n=4$ FSMFT yields currents within 2$\%$ of those computed from MC
simulations. The least-accurate $n=0$ FSMFT gives $J = q/(q+1)^{2}$
and is equivalent to previous treatments of a single defect
\cite{KOLODEFECT}.  Larger segments $n$ yield increasingly unwieldy
algebraic expression for $J_{1}(q)$.  The FSMFT (which is exact if $n=N$) is
asymptotically correct for $q\rightarrow (0,1)$ and is a systematic
expansion in $J = \sum_{i=1}^{\infty}a_{i}q^{i}$ about $q=0$. 
Coefficients up $a_{n+1}$ can be shown to be given correctly by
a $2n$-segment MFT, {\it i.e.}, for $n\geq 1, \,J\sim q -3q^{2}/2 + {\cal
O}(q^{3})$ rather than $J\sim q - 2q^{2}+{\cal O}(q^{3})$ predicted by
the $0$-segment MFT \cite{KOLODEFECT}.

The current $J$ can be further reduced upon adding successive,
contiguous defects.  The steady-state current across
$m$ equivalent, contiguous interior bottlenecks can be shown to have
the analytic expansion 

\begin{equation}
J_{m}(q\rightarrow 0) \sim \left({m+1 \over 4m-2}\right) q + {\cal O}(q^{2}).
\label{MCHAIN}
\end{equation}

\begin{figure}[h]
\begin{center}
\includegraphics[height=2.3in]{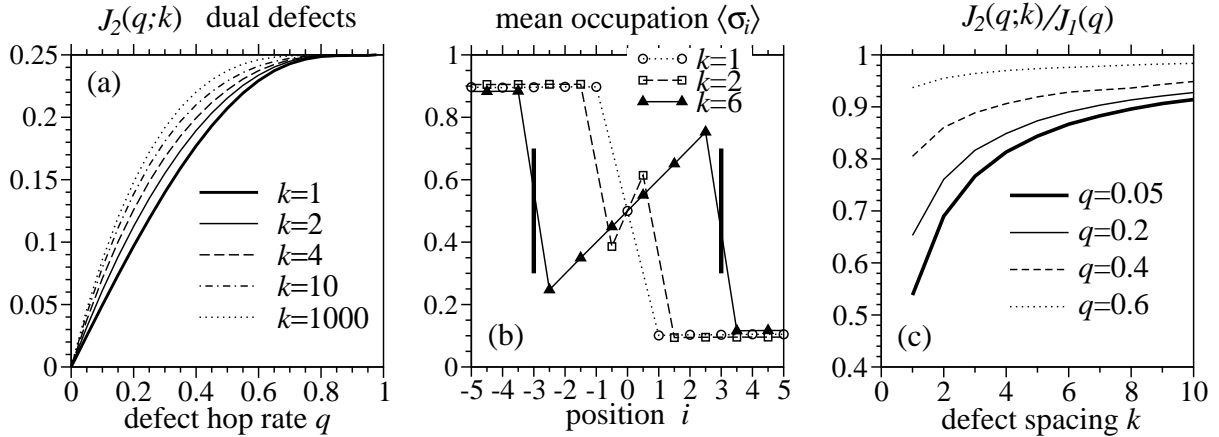}
\end{center}
\vspace{-4mm}
\caption{(a) Steady-state currents across a chain with two identical
defects in the interior spaces by $k$ lattice sites.  The current is
suppressed most when the defects are closely spaced. (b) The
density profiles centered at the midpoint 
between two defects ($q=0.15$) of various
spacings $k$.  The thick vertical bars denote the defect positions for
$k=6$. For larger $k$, density boundary layers heal, allowing particle
build-up behind the second barrier, enhancing the current. (c) The
dependence of the normalized steady-state current
$J_{2}(q;k)/J_{1}(q)$ as  a function of the defect separation.}
\label{double_mid2}
\end{figure}

\noindent As $m$ increases, the current is determined by the
rate-limiting segment which resembles a uniform chain of hopping rates
$q$ with relatively fast injection from, and extraction into, the remainder
of the chain with hopping rates $p_{i}=1$.  Thus, the current approaches
that of a long,  uniform chain with hopping rates $q$  in the maximal
current regime: $J =q/4$. Figure \ref{single_mid2}(b) plots  $J_{m}(q)/q$
for various $q$ as a function of $m$ defects. A 14-segment MFT was
used in to generate the results and was checked to provide high
numerical accuracy up to $m \approx 12$.  For all $q<1$, $J_{m}(q)/q$
approaches $1/4$ as $m\rightarrow N$, with the largest changes
occurring for small $m$ and $q$. The leading order approximation in
(\ref{MCHAIN}) is very accurate (compared to FSMFT or MC
simulations) for $q$ as large as $0.1$.

For strong bottlenecks with $q \lesssim 0.3$, the largest reduction in
$J_{m}(q)$ occurs as  $m=1 \rightarrow 2$. Therefore, one may
consider the effects of placing only two bottlenecks in the mRNA interior. 
Fig.  \ref{double_mid2}(a) shows the expected currents $J_{2}(q;k)$
across a chain containing two defects spaced $k$ sites apart.  Results for
$k \leq 10$  were computed using an $n=14$ FSMFT, while those for
$k>10$ were obtained from MC simulations.  The largest reduction in the
current occurs when two defects are spaced as closely as possible.  The
current $J_{2}(q;k\rightarrow \infty) \rightarrow J_{1}(q)$ eventually
approaches that for a {\it single} defect.  A finite number of multiple
defects, if spaced far apart, will not significantly decrease  $J$ relative to
the case of a single defect.  As $k$ increases, the density downstream
of the first defect recovers to the bulk value $\s_{-}$ before encountering
the second defect.  This behavior is clearly shown (using both FSMFT
and MC simulations)  in Fig. \ref{double_mid2}(b) for a pair of defects
($q=0.15$).  For two identical defects with small $q$  we find

\begin{equation}
J_{2}(q;k) \sim \left({k \over k+1}\right) q  + {\cal O}(q^{2}).
\end{equation}

\noindent Therefore, the current for two identical bottlenecks is at most
a factor of 2 smaller than that for a single defect. This maximal contrast
occurs when $q \rightarrow 0$ and when the two bottlenecks are
adjacent to each other. Fig. \ref{double_mid2}(c) plots the variation in
$J_{2}(q;k)/J_{1}(q)$ as a function of $k$ for various $q$.  

\begin{figure}[h]
\begin{center}
\includegraphics[height=2.3in]{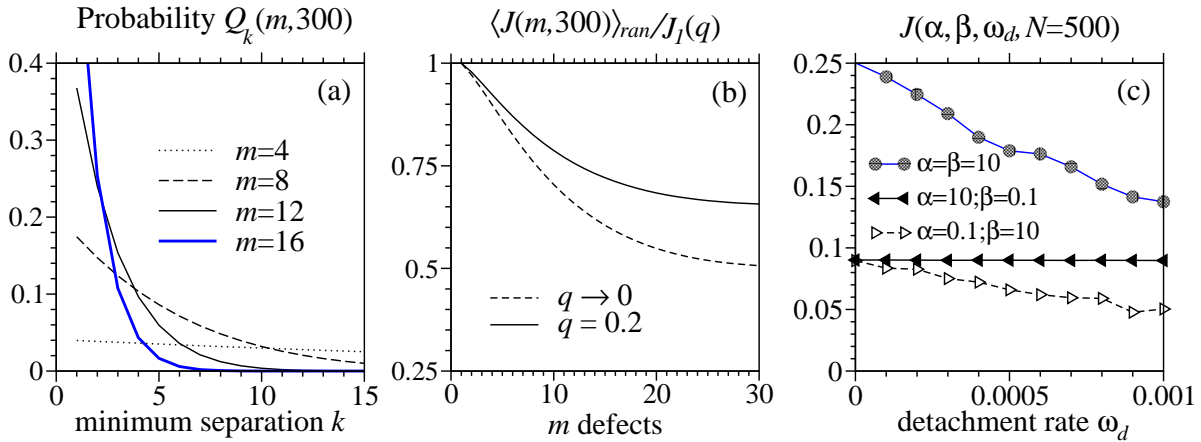}
\end{center}
\vspace{-3mm}
\caption{(a) The probability $Q_{k}(m,N)$ that the minimum separation of 
any two defects (of a total of $m$ on an $N$ site lattice section) is $k$
(Eqn. (\ref{QK}). (b) The
upper bound for the mean current of a lattice with $m$
defects randomly distributed within the central $N=300$ sites (Eqn. (\ref{Q})). 
This $J_{1}(q)$-normalized, 
randomness-averaged upper bound includes 
suppression only from pairwise defect configurations.
A lower bound as $m \rightarrow N-1$ 
is $\langle J(m\rightarrow N-1)\rangle_{ran} \rightarrow q/4$.
(c). The effects of a small, uniform detachment rate $\omega_{d}$ at each site.
Detachments suppress ribosome throughput only in the entry-limited
($\a=0.1, \b=10$) and maximal current ($\a=\b=10$) regimes.}
\label{double_mid3}
\end{figure}

Typically, a cell has a fixed number of rare codons with which to
implement different translation regulation strategies. From the
two-defect current $J_{2}$, we estimate the expected current for $m$ {\it
randomly} distributed bottlenecks.  The current will be reduced to a value
less than or equal to $J_{2}(q;k)$ only when two or more defects are
spaced within $k$ sites. The total number of ways $m$ defects can be
placed on $N-1$ interior sites, such that the minimum pair
spacing is $k$ is equivalent to the ``Tonk's gas'' 
partition function for $m$ particles 
of integer length $k$ distributed on a lattice with $N-1+k$ sites \cite{TG}:

\begin{equation}
Z_{k}(m,N) = {N-1-(m-1)(k-1) \choose m}.
\end{equation}

\noindent The
probability density that the minimum inter-defect spacing equals $k$ is thus

\begin{equation}
Q_{k}(m,N) = {(Z_{k}(m,N) - Z_{k+1}(m,N)) \over Z_{1}(m,N)}.
\label{QK}
\end{equation} 

\noindent The probability density $Q_{k}(m,300)$ is shown in Fig.
\ref{double_mid3}(a) for $m=4,8,12,16$. For small number of defects $m$, the
probability density is broad in $k$, indicating the large number of separations $k$
uniformly accessible to any of the few pairs. As $m$ increases, the likelihood of any
pair being close increases, pushing $Q_{k}(m,N)$ to peak at small $k$.

The maximum 
current for any configuration with a minimum defect spacing $k$ is $J_{2}(q;k)$.
Thus, an upper bound for the randomness-averaged current is 

\begin{equation}
\langle J \rangle_{ran} \leq\hspace{-8mm}
\sum_{k=1}^{\mbox{int}\{(N-1)/(m-1)\}}\hspace{-8mm}
Q_{k}(m,N)J_{2}(q;k).
\label{Q}
\end{equation}

\noindent This result will be very accurate if the defect density is low
enough that one can neglect the probability that more than two defects
each separated by $k$ sites form a single cluster, particularly for small
$k$. Although the most likely minimum defect spacing is $k=1$, at low defect
densities, the total probability of closely-spaced defects remains small
and the weight of $Q_{k}(m,N)$ at larger $k$ dominate the statistics
of $\langle J\rangle_{ran}$.  The disorder-averaged current
$\langle J(m,N=300) \rangle_{ran}$ (normalized by $J_{1}(q))$ is shown
in Fig.  \ref{double_mid3}(b).  For very small $m$, the current is
approximately that of a single defect.  The current is most
sensitive to the number of random defects at $m\approx 10$,
corresponding to $m/N\approx  0.03$, approximately the fraction of slow
codons observed {\it in vivo}. Nonetheless, the observation of enhanced,
nonrandom clustering suggests that other biological regulation
pathways exist and would yield currents measurably below the upper
bound in (\ref{Q}).  As $m \rightarrow N-1$, large clusters of defects
dominate that are not accounted for in $Q_{k}(m,N)$, and the current approaches
$q/4$. 

Finally, we consider the effects of a small uniform ribosome detachment at each site
with rate $\omega_{d}$. Monte-Carlo simulations show in Fig. \ref{double_mid3}(c)
that the steady-state current is diminished only when the TASEP is in what normally
would be the entry-limited (small $\a$, large $\b$) and maximal current (large $\a, \b$)
regimes.  When translation is initiation-limited, detachments occur for particles that
have already passed the rate-limiting initiation stage, so they no longer contribute to
the final current.  In the  maximal current regime, interior particles are also detaching,
thus reducing the number of particles that contribute to the current.  In the
exit-limited regime (large $\a$, small $\b$), detachments also  decrease the overall
ribosome density along most of the mRNA.  However, if $\omega_{d}$ is small (as in
Fig.  \ref{double_mid3}(c)), termination remains the rate-limiting step, ribosomes are
still able to pile up against the termination site,  and the final ribosome throughput $J
= \b \sigma_{N}$ remains relatively unchanged. In the termination rate-limited regime,
although the final current is not significantly suppressed by small $\omega_{d}$, the
high ribosomal density along the mRNA results in a large current {\it off} the mRNA
and may be metabolically wasteful.  


We have found that not only can a single defect directly inhibit elongation across it,
but a few bottlenecks, properly distributed, can further slow protein production by a
factor of $\sim 2-4$.\footnote{Ribosomes are structurally larger than nucleotide
triplets, occluding $w \sim 10$ codons. Although our FSMFT algorithm cannot be
easily adapted for large particles,  the steady-state properties of a $w > 1$ TASEP
\cite{LAKATOS,SHAW} are qualitatively the same as the standard TASEP were each
particle occupies exactly one lattice site, as verified by extensive MC simulations. 
However, the finite ribosome size prevents defects spaced $k\lesssim w$ codons
apart to influence each other.} Depending on the targeted biological outcome,
qualitatively distinct strategies can be employed.  Although maximal current reduction
is achieved by clustering defects as tightly as possible, successive addition beyond a
handful of defects does little to reduce the current. Defects which are all are spaced
more than a handful of sites apart will not reduce the throughput more than  a {\it
single} defect. These results are qualitatively consistent with a single, localized region
providing the rate limiting step(s) for translation.  Since initiation, which requires
assembly of numerous ribosome parts, is typically rate limiting, the observation of
slow codons  near the start codon \cite{CHEN} (forming the equivalent of two
closely-spaced defects near the start) suggests an expression-inhibiting, regulatory
role for these initiation-end bottlenecks.  Conversely,  a finite number of
well-separated defects at appropriate junctures provide pause points for local,
successive protein folding with {\it minimal} reduction in current. Additional processes
such as detachment, when combined with upstream or downstream bottlenecks can
also regulate ribosome throughput.

\vspace{5mm}

\noindent The authors thank L. Shaw and A. B. Kolomeisky for valuable
discussions.  TC acknowledges support from the NSF (DMS-0206733),
and the NIH (R01 AI41935).

\end{document}